\documentclass[useAMS,usenatbib,twocolumn]{mn2e}
\usepackage{xspace}
\usepackage{hyperref}
\usepackage{aas_macros}
\usepackage{url}
\usepackage{amsmath}
\usepackage{color}
\usepackage{amssymb}
\usepackage[capitalise]{cleveref}
\usepackage{graphicx}

\newcommand{\eobs}{\epsilon^{\mathrm{obs}}}

\newcommand{\Clobs}{\widetilde{C}_\ell}
\newcommand{\degsq}{\, \mathrm{deg}^2}

\newcommand{\tbf}{}

\newcommand{\om}{\Omega_\mr m}
\newcommand{\omb}{\Omega_\mr b}

\newcommand{\lcdm}{{\ensuremath{\Lambda\mathrm{CDM}}}\xspace}

\newcommand{\w}{w_0}
\newcommand{\wa}{w_a}
\renewcommand{\d}{{\rm d}}
\newcommand{\pd}{P_{\delta}}

\newcommand{\mr}{\mathrm}

\newcommand{\fsky}{f_{\rm sky}}
\newcommand{\for}{f_{\rm O-R}}

\newcommand{\jbca}{{Jodrell Bank Centre for Astrophysics, School of Physics \& Astronomy, The University of Manchester, Manchester M13 9PL, UK}}
\newcommand{\cosmosis}{\textsc{CosmoSIS}\xspace}
\newcommand{\des}{DES\xspace}
\newcommand{\euclid}{\textit{Euclid}-like\xspace}
\newcommand{\planck}{\textit{Planck}\xspace}
\title[SKA Weak Lensing I: Forecasts]{SKA Weak Lensing I: Cosmological Forecasts and the Power of Radio-Optical Cross-Correlations}
\author[Harrison et al.]{Ian Harrison\textsuperscript{\thanks{E-mail: ian.harrison-2@manchester.ac.uk}}, Stefano Camera, Joe Zuntz \& Michael L. Brown\\\jbca}

\begin{document}
%\date{Accepted 1988 December 15. Received 1988 December 14; in original form 1988 October 11}

\pagerange{\pageref{firstpage}--\pageref{lastpage}} \pubyear{2016}

\maketitle

\label{firstpage}

\begin{abstract}
We construct forecasts for cosmological parameter constraints from weak gravitational lensing surveys involving the Square Kilometre Array (SKA). Considering matter content, dark energy and modified gravity parameters, we show that the first phase of the SKA (SKA1) can be competitive with other Stage III experiments such as the Dark Energy Survey (DES) and that the full SKA (SKA2) can potentially form tighter constraints than Stage IV optical weak lensing experiments, such as those that will be conducted with LSST, WFIRST-AFTA or \textit{Euclid}-like facilities. Using weak lensing alone, going from SKA1 to SKA2 represents improvements by factors of $\sim10$ in matter, $\sim10$ in dark energy and $\sim5$ in modified gravity parameters. We also show, for the first time, the powerful result that comparably tight constraints (within $\sim5\%$) for both Stage III and Stage IV experiments, can be gained from cross-correlating shear maps between the optical and radio wavebands, a process which can also eliminate a number of potential sources of systematic errors which can otherwise limit the utility of weak lensing cosmology.
\end{abstract}

\section{Introduction}
\label{sec:introduction}
Mapping the cosmic shear signal with weak gravitational lensing has long been regarded as an excellent probe of cosmology \citep[see e.g.][for a recent review]{2015RPPh...78h6901K}. In particular, future weak lensing measurements are one of the most promising observables for constraining the history of the growth of cosmic structure (and the physics which caused it) through direct sensitivity to the total mass along a line of sight \citep[e.g.][]{2013PhR...530...87W}. 

From early detections \citep{2000MNRAS.318..625B, 2000Natur.405..143W,2000A&A...358...30V,2000astro.ph..3338K}, progress has been made to the point whereby current experiments \citep{2013MNRAS.432.2433H, 2015arXiv151003962J, 2015arXiv150705552T} are able to provide matter contents and dark energy constraints comparable with the best available from other probes such as the Cosmic Microwave Background (CMB, \citealt{2015arXiv150201589P}) and galaxy clustering \citep{2012PhRvD..86j3518P, 2013A&A...557A..54D, 2014MNRAS.441...24A}. As the depth and sky area of these and future experiments increases, uncertainties on these constraints will begin to become dominated by the numerous systematic effects which come into play when turning the raw astronomical data into shear maps and subsequent parameter confidence regions. These systematics include (but are not limited to) telescope systematics, galaxy intrinsic alignments \citep[see e.g.][]{2015SSRv..193....1J}, image analysis algorithm errors and uncertainties associated with modelling the non-linearity of matter clustering on small physical scales. 

In this paper we will consider in particular the promise of future weak lensing experiments involving the Square Kilometre Array (SKA)\footnote{\url{http://www.skatelescope.org}} radio interferometer telescope, both alone and in cross-correlation with representative optical weak lensing surveys. The SKA has unique value by itself, the exact extent of which will depend on the properties of the faint radio source population which will be probed by surveys with SKA pathfinders and precursors. In an ideal scenario, the properties of this population will contain a long-tailed source redshift distributions, expected for the star-forming galaxy (SFG) population that will dominate the SKA surveys, and add unique additional information on the lensing shear signal from radio polarisation and resolved spectral line observations \citep[see][for a summary]{2015aska.confE..23B}. Even without the addition of more information, extra advantages can also be gained by cross-correlating the shear maps produced from SKA data with shear maps generated by other experiments in different wavebands, as recently demonstrated by \cite{2015arXiv150705977D}. In this procedure, any spurious shear generated by systematics which are uncorrelated between the wavebands should be instantly eliminated \citep[e.g.][]{2010MNRAS.401.2572P}. In particular, contamination from an incorrectly deconvolved spatially varying Point Spread Function (PSF) and errors from algorithms used to measure the shapes of individual galaxies to infer the shear should be uncorrelated between the different experiments. When measuring an observed shear map $\widetilde{\gamma}$ made in waveband ${X}$, the observed signal receives contributions from the true gravitational shearing $\gamma$ (which is achromatic and identical in both wavebands), the intrinsic shape of the galaxy $\gamma^{\rm int}$ and spurious shear from incorrectly deconvolved PSF or shape measurement error $\gamma^{\rm sys}$. The cross-correlation of shear maps in different wavebands then has terms:
\begin{equation}
\begin{split}
\langle\widetilde{\gamma}_{X}\widetilde{\gamma}_{Y}\rangle =
\langle \gamma \gamma \rangle +& \langle \gamma^{\rm int}_{X} \gamma \rangle + \langle \gamma^{\rm int}_{Y} \gamma \rangle \\ &\quad + \langle \gamma^{\rm int}_{X} \gamma^{\rm int}_{Y} \rangle +  \langle \gamma^{\rm sys}_{X} \gamma^{\rm sys}_{Y} \rangle.
\end{split}
\end{equation}
The first term is the cosmological signal that we are interested in, the following three terms are contaminating `intrinsic alignment' terms \citep[see][for a recent review]{2015SSRv..193....1J, 2015SSRv..193...67K, 2015SSRv..193..139K} and the final term is a systematics term (we have ignored terms correlating systematics with signals on the sky). Any contributions to these systematics terms which are uncorrelated between different experiments and wavebands will be suppressed by the cross-correlation, greatly increasing the robustness of cosmological constraints. \tbf{If polarised and neutral hydrogen (HI) 21 cm line emission fractions from high redshift sources prove to be high enough, radio weak lensing experiments can also provide useful information} on intrinsic alignment systematics through polarisation \citep{2011MNRAS.410.2057B} and rotational velocity information \citep{2002ApJ...570L..51B, 2006ApJ...650L..21M}, though we do not consider such approaches in these forecasts. Instead, we consider what can be achieved with `vanilla' SKA weak lensing surveys in which cosmological information come from forming shear power spectra from measured galaxy ellipticities, just as in typical optical experiments. Adopting the survey categorisation scheme of the Dark Energy Task Force \citep[DETF,][]{2006astro.ph..9591A}, we will show that surveys conducted with the first phase of the SKA (SKA1) will be competitive with `Stage III' optical weak lensing surveys such as DES\footnote{\url{http://www.darkenergysurvey.org}}, KiDS\footnote{\url{http://kids.strw.leidenuniv.nl}} and HSC\footnote{\url{http://subarutelescope.org/Projects/HSC}}, and that full SKA (SKA2) weak lensing surveys can provide `Stage IV' constraints similar to those achievable with the weak lensing components of the \textit{Euclid}\footnote{\url{http://euclid-ec.org}}, WFIRST-AFTA\footnote{\url{http://wfirst.gsfc.nasa.gov}} and LSST\footnote{\url{http://www.lsst.org}} surveys. We will also show that constraints obtained from cross-power spectra measured between shear maps made in different wavebands will provide measurements which are still just as tight as each experiment by itself, but should be free of any wavelength dependent systematics.

Here we make forecasts using simple prescriptions for the noise spectra and covariance matrices within a weak lensing experiment, and choose a fiducial experimental configuration for the SKA weak lensing surveys. In a companion paper \citep[][hereafter Paper II]{bonaldi2016} we construct a sophisticated simulation pipeline to produce mock weak lensing catalogues for future SKA surveys which we also process through a tomographic weak lensing power spectrum analysis. We then use this pipeline to explore the optimal instrumental configuration for performing SKA weak lensing surveys in the presence of real-world effects such as signal-to-noise dependent shape measurement errors, realistic distributions in galaxy sizes, fluxes and redshifts and ionospheric distortions.

The outline of this paper is as follows. We first provide a brief review of radio weak lensing in \cref{sec:wl}. In \cref{sec:experiments} we then describe the experimental surveys considered for the forecasts and describe our methodology for construction of cross-experiment shear power spectra. In \cref{sec:forecasts} we describe the methods used in producing our forecasts. Then, in \cref{sec:results} we show results for cosmological parameter constraints using SKA, Stage III optical (\des), Stage IV optical (\euclid) and cross-correlations, demonstrating the power of using optical and radio experiments together. Finally in \cref{sec:conclusions} we discuss these results and conclude.

\section{Weak Lensing Cosmology}
\label{sec:wl}
We refer the reader to \cite{2001PhR...340..291B} for a comprehensive overview of weak lensing cosmology, which we will briefly introduce here. Weak lensing analyses typically involve the measurement of the individual shapes of large numbers of galaxies on the sky. For a large number density of sources in a single patch of sky, the estimated change in shape due to the cosmic shear along the line of sight to that patch ($\hat{\gamma}$) can be estimated by taking a simple average over the observed ellipticity of the galaxies ($\eobs$), assuming that the intrinsic shapes before shearing are uncorrelated:
\begin{equation}
\label{eqn:gamma_hat}
\hat{\gamma} = \frac{1}{N}\sum_{i=1}^{N}\eobs_{i}.
\end{equation}
The two-point statistics of this observed shear field, such as the power spectrum $\Clobs$, can then be related to the underlying matter power spectrum $\pd$, which can be predicted theoretically for different cosmological models. For sources confined to a thin shell in redshift, the $\Clobs$ are sensitive to the integrated matter power spectrum out to this redshift. In practice, sources are distributed across a range of redshifts $\d n_{\rm gal}/\d z$ (which is in turn affected by imprecise knowledge of the redshifts of individual sources) and extra information is gained about the growth of structures along the line of sight by constructing the auto- and cross-power spectra of shear maps made using sources divided into different tomographic redshift bins.

The full relation for the power spectrum between two different tomographic bins $i,j$ is given by \citep{2001PhR...340..291B}:
\begin{equation}
\label{eqn:limber}
C ^{ij} _{\ell} = \frac{9H_0^4 \om^2}{4c^4} \int_0^{\chi_\mr h} 
\mr d \chi \, \frac{g^{i}(\chi) g^{j}(\chi)}{a^2(\chi)} \pd \left(\frac{\ell}{f_K(\chi)},\chi \right) \,.
\end{equation}
Here, $H_0$ is the Hubble constant, $\om$
is the (total) matter density, $c$ is the speed of light, $a(\chi)$
is the scale factor of the Universe at co-moving distance $\chi$,
$f_K(\chi)$ is the angular diameter distance (given simply by
$f_K(\chi) = \chi$ in a flat Universe), $\pd(k, \chi)$ is the
matter power spectrum and the functions $g^{i}(\chi)$ are the lensing
kernels for the redshift bins in question. The lensing kernels are given by:
\begin{equation}
g^{i}(\chi) = \int_\chi^{\chi_{\mr h}} \mr d \chi' n_{i} (\chi') \frac{f_K (\chi'-\chi)}{f_K (\chi')} \,.
\end{equation}
The number density distributions $n_{i} (\chi)$ give the normalised number of galaxies with radial co-ordinate $\chi$ in this tomographic bin. For single experiment weak lensing cosmology, the $i,j$ label different tomographic redshift bins and the uncertainty on the power spectrum depends on $n_{\rm gal}$, the number density of detected galaxies on the sky and $\sigma_{g}$, the variance of the distribution of galaxy ellipticities (or `shape noise'). We will generalise these measurement and noise terms to include cross-experiment power spectra in \cref{sec:cross-spectra}.
\subsection{Cosmological Parameters}
In this paper we will consider the ability of weak lensing experiments to measure a base six-parameter \lcdm model and two well-motivated extensions: dynamical dark energy and a phenomenological modification to Einstein's gravity. We note that these choices are merely common parametrisations of these extensions and are not specifically tailored to the strengths of SKA weak lensing. Different parametrisations (for example, non-parametric dark energy equation of state reconstruction which equally weights information at all redshifts) may more optimally use the information from these experiments for model selection, but are not considered here.
\subsubsection{Base \lcdm}
For our base cosmology we consider six parameters: total matter content $\om$, baryonic matter content $\omb$, amplitude of matter fluctuations $\sigma_8$, Hubble expansion parameter $h_0$, scalar fluctuation spectral index $n_s$ and reionisation optical depth $\tau$. Unless otherwise stated, all constraints presented are marginalised over the first five of these parameters (with $\tau$ kept fixed) with central values of $\boldsymbol\vartheta_\lcdm = \lbrace \om, \omb, \sigma_8, h_0, n_s \rbrace = \lbrace 0.3, 0.04, 0.8, 0.72, 0.96 \rbrace $. Weak lensing is highly effective at probing the overall amplitude of the matter power spectrum, which depends on a degenerate combination of the total matter $\om$ and clustering strength $\sigma_8$; we will therefore present constraints in these two parameters only.
\subsubsection{Dark Energy}
As one extension to $\Lambda$CDM, we will consider measuring the parameters in a simple model of evolving dark energy where the equation of state $w$ evolves as a linear function of the scale factor $a$ (known as the Chevallier-Polarski-Linder parameterisation, see \citealt{Chevallier:2000qy} and \citealt{Linder:2002et}):
\begin{equation}
w(a) = w_0 + w_a(1 - a).
\end{equation}
This model represents the first order term in a Taylor expansion of a generally evolving equation of state. We consider these parameters in $\boldsymbol\vartheta_w = \boldsymbol\vartheta_\lcdm + \lbrace w_0, w_a \rbrace$.
\subsubsection{Modified Gravity}
We also consider modifications to gravity as parametrised in \cite{2011PhRvD..84l3001D, 2015PhRvD..92b3003D}. In General Relativity, from the perturbed Friedmann-Lemaitre-Robertson-Walker (FLRW) metric in the conformal Newtonian gauge:
\begin{equation}
ds^2 = a^2(\eta) \left[ -(1 + 2\Psi)d\eta^2 + (1- 2\Phi)dx^adx_a \right],
\end{equation}
we define the Newtonian gravitational potential $\Psi$ felt by matter and the lensing potential $\Phi$ which is also felt by relativistic particles. We now define modified gravity parameters $Q_0$, which modifies the potential $\Phi$ in the relativistic Poisson equation:
\begin{equation}
k^2 \Phi = -4\pi G a^2 \rho \Delta Q_0
\end{equation}
and the gravitational slip $R$ which, in the case of anisotropic stress, gives the ratio between the two potentials:
\begin{equation}
R = \frac{\Psi}{\Phi}.
\end{equation}
As $R$ is degenerate with $Q_0$ it is convenient to define the derived parameter $\Sigma_0 = Q_0(1+R)/2$ and our constraints are given in terms of this. Weak lensing probes the sum of potentials $\Phi + \Psi$ and is hence extremely effective at constraining $\Sigma_0$ but much less sensitive to $Q_0$. Combination with probes for which the opposite is true (i.e. which are sensitive to the Newtonian potential), such as redshift space distortions, is then capable of breaking the degeneracy inherent in each probe individually \citep[see e.g.][]{2013MNRAS.429.2249S, 2015PhRvD..91h3504L}. We consider these parameters in $\boldsymbol\vartheta_{mg} = \boldsymbol\vartheta_\lcdm + \lbrace \Sigma_0, Q_0 \rbrace$.

\subsection{Weak Lensing Systematics}
Whilst the statistical error on a weak lensing measurement of a cosmological parameter can be beaten down through increasing the number density of galaxies $n_{\rm gal}$ with measured shapes on the sky (or by selecting a population with a smaller intrinsic shape dispersion $\sigma_{g}$), forthcoming Stage III and Stage IV experiments will begin to enter the regime where the contribution from systematic errors on shear measurement will become comparable to, and larger than, the statistical noise. Here we provide a brief overview of many (although not all) of these systematics, whereas a more detailed analysis of their effects and ways to overcome them will be provided in a companion paper \citep[][hereafter Paper III]{camera2016}.
\begin{itemize}
\item PSF uncertainties. The light from all sources used in weak lensing is convolved with the telescope point spread function. This convolution will induce changes in the size and ellipticity of the apparent galaxy shape in the image data, and must be accounted for when estimating the true observed ellipticity. Typically, a model is created for the PSF which is then deconvolved during shear measurement. For ground-based optical experiments, the primary systematic is residual, un-modelled PSF shape distortions due to instabilities in the atmosphere above the telescope (i.e. seeing). For space-based telescopes the atmosphere is not a consideration, but other effects from detectors and telescope optics can still create an anisotropic and time-varying PSF. \tbf{In addition, the deterministic nature of the changes in interferometer dirty beam shape with observing frequency may potentially avoid issues with shear bias from colour gradients in source galaxies \citep[see e.g.][for a full description of the problem]{2012MNRAS.421.1385V}. However, care will need to be taken to ensure the primary beam of each antenna is well-characterised enough to avoid the return of shear biases originating from the beam.}
\item Shear measurement uncertainties \citep[see][and references therein for an overview]{2014ApJS..212....5M}. Using the observed galaxy ellipticity as a shear estimator as in \cref{eqn:gamma_hat} depends on having a reliable, unbiased estimator of the ellipticity. Whilst in the noise-free case, $\epsilon$ can be defined as a simple function of the quadrupole moments of the image, significant complications arise whenever noise is present as the un-weighted quadrupoles will diverge. In general, maximum likelihood estimators for ellipticity will become increasingly biased at lower signal-to-noise ratios (as ellipticity is a ratio of quadrupole moments), and so must be calibrated \citep[e.g.][]{2012MNRAS.425.1951R}. Shear estimators which measure $\epsilon$ using parametrised models with elliptical isophotes also suffer from `model bias' caused by under-fitting of real galaxy intensity profiles \citep{2010MNRAS.404..458V}. Accounting for these biases correctly, through either explicit calibration or application of correct Bayesian priors, is a major step in the analysis pipeline for most surveys and requires sophisticated, large scale simulations which correctly reflect the observations.
\item Intrinsic Alignment (IA) contamination. A key assumption in \cref{eqn:gamma_hat} is that intrinsic galaxy shapes are uncorrelated and so any coherent shape must be due to cosmic shear. However, in reality there are two other astrophysical effects which contaminate the shear signal. Galaxies which are nearby on the sky form within the same large scale structure environment as one another, creating spurious `II' (Intrinsic-Intrinsic) correlations. In addition, galaxies which are local in redshift to an overdensity will develop intrinsic shapes in anti-correlation with the shearing of background galaxies by that same overdensity -- the `GI' (Gravitational-Intrinsic) alignment. Typically, these alignments can be \tbf{mitigated} through modelling their effect on the power spectrum, or discounting galaxies which are expected to be most affected (such as close pairs on the sky or redder galaxies). An overiew of IA effects can be found in \citet{2015SSRv..193....1J}, \citet{2015SSRv..193...67K} and \citet{2015SSRv..193..139K}.
\item Non-linear evolution and baryonic feedback effects. Cosmology with cosmic shear relies on the comparison between an observed shear power spectrum and a theoretically predicted one. However, outside of the regime of linear evolution of large scale structures (i.e. on smaller scales $k \gtrsim 0.2 h \, \mathrm{Mpc}^{-1} $), a variety of physical effects will affect the shape of this power spectrum in uncertain ways which are possibly degenerate with changes in cosmological parameters \citep[e.g.][]{2005APh....23..369H}.
\item Redshift uncertainty estimation. Placing sources into tomographic bins usually requires an estimate of the source's redshift from a small number of broad photometric bands. Significant biases may arise due to insufficient freedom in Spectral Energy Distribution (SED) templates, incorrect spectroscopic calibration and noisy data. For a discussion of these issues see \cite{2015arXiv150705909B} and references therein.
\end{itemize}

\subsection{Radio Weak Lensing}
Performing weak lensing experiments in the radio band offers a number of potential advantages compared to using optical telescopes alone. In addition to opening the door to powerful cross-correlation techniques (which we consider in more detail in the following subsection), the radio band has the potential to bring unique added value to this area of cosmology by way of new approaches to measuring the weak lensing signal using polarisation and rotational velocity observations. Here we summarise the key benefits that radio weak lensing experiments can offer and highlight some of the challenges which need to be met. We refer the reader to \cite{2015aska.confE..23B} for more information.   

\begin{itemize}
\item Weak lensing surveys conducted with radio telescopes are, in principle, much less susceptible to instrumental systematic effects associated with residual PSF anisotropies. \tbf{The image-plane PSF (or `dirty beam') is set by the baseline distribution and time and frequency sampling of the telescope, all of which are deterministic and known to the observer and may be controlled. An anisotropic PSF can mimic the sought-after cosmic shear signal and are one of the most worrisome systematic effects in optical lensing analyses.} Whilst the turbulent ionosphere can cause similar effects in the radio, these effects scale strongly with frequency, meaning at the high frequency considered here ($1.355\,\mathrm{GHz}$, see Paper II for a full discussion) this is less of a concern for radio weak lensing.
\tbf{\item However, whilst the dirty beam is precisely known and highly deterministic, the incomplete sampling of the Fourier plane by the finite number of interferometer baselines leads to significant sidelobes which may extend across the entire visible sky. Deconvolving this PSF then becomes a complicated non-local problem as flux from widely-separated sources is mixed together and traditional methods (such as the CLEAN algorithm, \citealt{1974A&AS...15..417H}) have been shown to be inadequate for preserving morphology to the degree necessary for weak lensing.}
\item The SFGs which are expected to dominate the deep, wide-field surveys to be undertaken with the SKA are also expected to be widely distributed in redshift space \citep[see][and Paper II]{wilman08}. In particular, a high-redshift tail of significant numbers of such galaxies, extending beyond $z \sim 1$ would provide an additional high-$z$ bin to what is already accessible with optical surveys. \tbf{See the end of \cref{sec:fisher} for a demonstration of the increase in cosmological constraining power from the inclusion of these high-redshift sources. The details of the flux and size distributions of this population are still somewhat uncertain (see Paper II for a full discussion) and will benefit from the efforts of SKA precursor and pathfinder surveys.}
\item The orientation of the integrated polarised emission from SFGs is not altered by gravitational lensing. If the polarisation orientation is also related to the intrinsic structure of the host galaxy then this provides a powerful method for calibrating and controlling intrinsic galaxy alignments which are the most worrying astrophysical systematic effect for precision weak lensing studies \citep{2011MNRAS.410.2057B, 2015MNRAS.451..383W}. \tbf{Again, the polarisation fraction and angle of scatter between position and polarisation angle is currently subject to much uncertainty and have currently only been tested on small low-redshift samples \citep{2009ApJ...693.1392S}. This result may not preserve in the high-redshift SFGs we are interested in here, but will become better informed by other surveys leading up to the SKA.}
\item Much like the polarisation technique, observations of the rotation axis of disk galaxies also provides information on the original (un-lensed) galaxy shape \citep{2002ApJ...570L..51B, 2006ApJ...650L..21M, huff2013}. Such rotation axis measurements \tbf{may be available} for significant numbers of galaxies with future SKA surveys through resolved 21 cm HI line observations.
\item HI line observations also provide an opportunity to obtain spectroscopic redshifts for sources used in weak lensing surveys \cite[e.g.][]{2015MNRAS.450.2251Y}, greatly improving the tomographic reconstruction \tbf{for the sources for which spectra are available. For SKA1 this will be a relatively small fraction of sources ($\sim10\%$) at low redshifts (which are less useful for gravitational lensing) but this will improve significantly for SKA2.}
\item Because Galactic radio emission at relevant frequencies is smooth, it is `resolved out' by radio interferometers. This means that radio surveys have access to more of the sky than experiments in other wavebands, which cannot see through the Galaxy because of dust obscuration effects. 
\end{itemize}

A detection of a weak lensing signal in radio data was first made by \cite{2004ApJ...617..794C} in a shallow, wide-area survey. More recently \cite{2015arXiv150705977D} have made a measurement in cross-correlation with optical data, and the SuperCLASS\footnote{\url{http://www.e-merlin.ac.uk/legacy/projects/superclass.html}} survey is currently gathering data with the express purpose of pushing forward radio weak lensing techniques.

\subsection{Shear Cross-Correlations}
\label{sec:cross-spectra}
Whilst radio weak lensing surveys have worth in themselves, as discussed above, combining shear maps made at different observational wavelengths has further potential to remove systematics which can otherwise overwhelm the cosmological signal. Here we construct a formalism for forecasting the precision with which cross-correlation power spectra can be measured from shear maps obtained from two different experiments $X,Y$, which may be in different wavebands. We may still split sources in each experiment into different redshift bins $i,j$, giving the cross power spectra:
\begin{equation}
\label{eqn:limber_cross}
C ^{X_{i}Y_{j}} _\ell = \frac{9H_0^4 \om^2}{4c^4} \int_0^{\chi_\mr h} 
\mr d \chi \, \frac{g^{X_i}(\chi) g^{Y_j}(\chi)}{a^2(\chi)} \pd \left(\frac{\ell}{f_K(\chi)},\chi \right) \,.
\end{equation}
Here the bins can be defined differently for each experiment, taking advantage of e.g. higher median redshift distributions or better measured photometric redshifts in one or the other of the two experiments. 

When observed, each power spectrum also includes a noise power spectrum from the galaxy sample:
\begin{equation}
\Clobs^{X_i Y_j} = C_\ell^{X_i Y_j} + \mathcal{N}_\ell^{X_i Y_j}.
\end{equation}
The noise is a function of the number density of galaxies in each experiment individually $n_{\rm gal}^{X_i}, n_{\rm gal}^{Y_j}$, the number of objects which are common to both experiments $n_{\rm gal}^{X_iY_j}$ and the covariance of galaxy shapes between the two experiments and redshift bins $\mathrm{cov}(\epsilon_{X_i}, \epsilon_{Y_j})$. Note that this final term $\mathrm{cov}(\epsilon_{X_i}, \epsilon_{Y_j})$ is in general a function of both waveband $X,Y$ and redshift bin $i,j$, describing how galaxy shapes are correlated between the two wavebands and how this correlation evolves with redshift. We can then write the expression for the noise on an observed shear power spectrum:
\begin{align}
\mathcal{N}_\ell^{X_i Y_j} &= \frac{1}{n_{\rm gal}^{X_i}n_{\rm gal}^{Y_j}}\langle \sum_{\alpha \in X_i}\epsilon_\alpha  \sum_{\beta \in Y_j}\epsilon_\beta \rangle \nonumber \\
&= \frac{n_{\rm gal}^{X_i Y_j}}{n_{\rm gal}^{X_i}n_{\rm gal}^{Y_j}}\mathrm{cov}(\epsilon_{X_i}, \epsilon_{Y_j}).
\end{align}
For correlations between redshift bins in the same experiment this reduces to the familiar shape noise term \citep[e.g.][]{2004PhRvD..70d3009H}:
\begin{equation}
\label{eqn:autonoise}
\mathcal{N}_\ell^{ij} = \delta^{ij}\frac{\sigma^2_{g_i}}{n^{i}_{\rm gal}}.
\end{equation}
If we make the simplifying assumption that for cross-experiment correlations, where redshift bins overlap, both experiments probe the same populations of galaxies which have the same shape and shape variance in both wavebands and across all redshift bins, the noise term becomes:
\begin{equation}
\label{eqn:simplenoise}
\mathcal{N}_\ell^{X_i Y_j} = \frac{n_{\rm gal}^{X_i Y_j}}{n_{\rm gal}^{X_i}n_{\rm gal}^{Y_j}}\sigma^2_{g}.
\end{equation}
% The extent to which this extreme case is true or false is the subject of ongoing investigation. To date there is conflicting evidence, with \cite{2009MNRAS.399.1888B} finding strong correlations between shapes in SDSS optical and FIRST radio data and \cite{2010MNRAS.401.2572P} finding little correlation in shapes of galaxies detected in the Hubble Deep Field-North (HDF-N) by the Hubble Space Telescope and the MERLIN radio interferometer. As pointed out by \cite{2010MNRAS.401.2572P}, this discrepancy could be due to selection effects --- the \cite{2009MNRAS.399.1888B} study predominantly selected AGN-type objects whereas the population probed by \cite{2010MNRAS.401.2572P} were mostly SFGs. However, the number statistics in the \cite{2010MNRAS.401.2572P} study were small and the situation is currently far from resolved. 

% We note that the galaxy population expected in future SKA surveys will be most like the population studied by \cite{2010MNRAS.401.2572P}. If the lack of optical-radio shape correlations that they found persists in the SKA population as a whole then \cref{eqn:simplenoise} will in fact \emph{over-estimate} the noise contribution to the cross-correlation measurement. In this sense our forecasts for the cross-correlations are conservative.

\tbf{Here, for the two sets of tomographic redshift bins for each experiment we consider the fraction of sources which may be expected to appear in both the radio and optical shape catalogues. In reality, this overlap will be between a deep optical sample and a deep radio sample of SFGs on a wide area. Data sets with this combination of area coverage and depth do not as yet exist, but useful information can be gained from some shallower or narrower archival surveys. Here we consider the large but shallow SDSS-DR10 optical catalogue \citep{2014ApJS..211...17A} and the FIRST radio catalogue \citep{1995ApJ...450..559B, 2004ApJ...617..794C}; and deep but narrow observations of the COSMOS field using the Hubble Space Telescope \citep{2010ApJ...708..202M} in the optical and VLA in the radio \citep{2010ApJS..188..384S}. The SDSS-FIRST overlap region contains a significant part ($\sim10,000 \, \degsq$) of the northern sky, but the radio catalogue is shallow (a $10 \, \sigma$ detection limit of $1.5 \, m$Jy). The COSMOS overlap survey is deep (a $10 \, \sigma$ detection limit of $0.28 \, m$Jy) but covers only 1 $\degsq$. These data sets appear to indicate that matching fractions are low ($<10\%$) and do not evolve significantly with redshift. In addition, the optical and radio weak lensing samples constructed by \cite{2010MNRAS.401.2572P} in an $8.5' \times 8.5'$ field in the HDF-N region contain a $4.2\%$ matching fraction across all redshifts.}
% \begin{figure}
% \includegraphics[width=\columnwidth]{both_redshift_matches.png}
% \caption{Optical-radio matching fractions.}
% \label{fig:matching_fractions}
% \end{figure}

\begin{figure}
\includegraphics[width=0.5\textwidth]{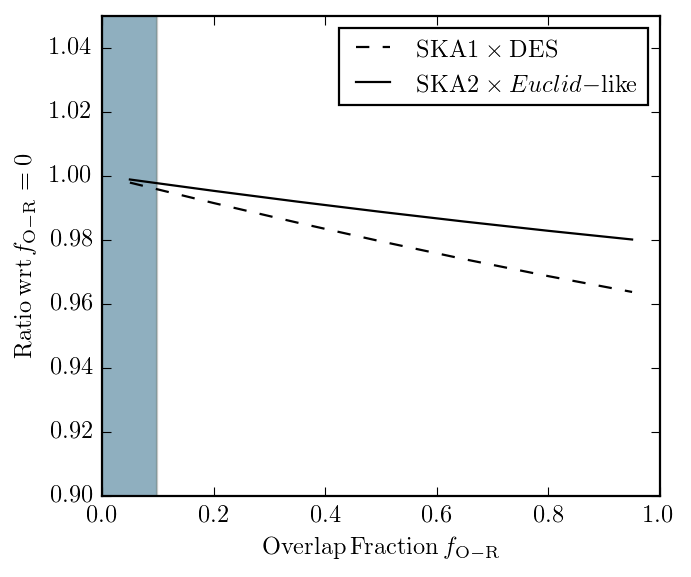}
\caption{\tbf{Ratio with respect to the case with no radio-optical matching fractions ($\for=0$) for dark energy FoMs as a function of $\for$ for the cross-correlation between Stage III (dashed line) and Stage IV (solid line) experiments. The shaded regions shows the range of values for $\for$ for the data sets discussed in the text.}}\label{fig:FoM_fOR}
\end{figure}

To investigate how much a non-vanishing radio-optical matching fraction could degrade the radio-optical cross-correlation constraining power for cosmology, we proceed as follows. We introduce a parameter $\for\in[0,1]$ quantifying the number of sources that appears in both the radio and the optical/near-infrared catalogues for a given combination of tomographic bins. In other words, we keep $n_{\rm gal}^{X_i Y_j}$ fixed to the amount of sources present in the overlap between to given radio-optical bin pairs $X_i-Y_j$. We then multiply this quantity by $\for$ and perform a Fisher matrix analysis letting $\for$ vary continuously between 0 and 1 (but identically across all redshift bins). \cref{fig:FoM_fOR} illustrates the degradation of the Dark Energy Task Force Figure of Merit (DETF FoM -- the inverse area of a Fisher ellipse in the $w_0$-$w_a$ plane, see \citealt{2006astro.ph..9591A} and \cref{eqn:de_fom}) -- as the fraction of matching radio-optical sources, $\for$, increases (note that for simplicity we assume $\mathrm{cov}(\epsilon_{X_i}, \epsilon_{Y_j})=\sigma_\epsilon^2$). We show the ratio between the DETF FoM for a non-vanishing radio-optical matching fraction $\for$ and the same quantity for $\for=0$. It is easy to see that even if 100\% of the sources appeared in both catalogues, the degradation of the dark energy FoM would be $<5\%$ for Stage III cosmic shear surveys, and even lower for Stage IV experiments. If we then consider the available data as described previously in this section, the range of values of $\for$ for which are indicated by the shaded area, we may see the minimal impact of realistic noise terms on the cross-correlation power spectra.

\tbf{In order to account for this in the following forecasts we consider the regime where overlap fractions are high and photometric redshifts are provided for the $85\%$ and $50\%$ of sources which do not have spectroscopic HI 21cm line redshifts in the case of SKA1 and SKA2 respectively (as described in \cref{tab:experiments}). However, as mentioned in \cref{sec:cross-correlations}, it may be possible for radio surveys alone to provide significantly more redshifts than those from only high-significance HI detections.}

In the regime where systematics are controlled, the maximum amount of information is available by using both cross and auto-experiment power spectra. For a data vector consisting of both:
\begin{equation}
\widetilde{\mathbf{d}} = 
\begin{pmatrix}
\Clobs^{XX} \\
\Clobs^{XY} \\
\Clobs^{YY}
\end{pmatrix},
\end{equation}
we can also write the covariance matrix between two bins in different experiments (now suppressing the $i,j$ for clarity and with $\nu =\delta_{\ell \ell'}/(2\ell + 1)\fsky$):
\begin{align}
\widetilde{\mathbf{\Gamma}}_{\ell\ell'} &= \\
& \nu \begin{pmatrix}
2(\Clobs^{XX})^2 & 2\Clobs^{XX}\Clobs^{XY} & 2(\Clobs^{XY})^2 \\
2\Clobs^{XX}\Clobs^{XY} & (\Clobs^{XY})^2 + \Clobs^{XX}\Clobs^{YY} & 2\Clobs^{XY}\Clobs^{YY} \\
2(\Clobs^{XY})^2 & 2\Clobs^{XY}\Clobs^{YY} & 2(\Clobs^{YY})^2
\end{pmatrix}, \nonumber
\end{align}
making the simplifying assumption that different $\ell$ modes are uncorrelated and hence the covariance matrix is diagonal in $\ell-\ell'$. However, here we are interested in forecasting constraints which can be gained which are free of systematics caused by e.g. incorrect PSF deconvolution within an experiment and so consider only cross-experiment spectra (as such systematics will be uncorrelated between the two experiments), giving data vector:
\begin{equation}
\label{eqn:simple_data}
\widetilde{\mathbf{d}} = 
\begin{pmatrix}
\Clobs^{XY}
\end{pmatrix},
\end{equation}
and covariance matrix:
\begin{equation}
\label{eqn:simple_cov}
\widetilde{\mathbf{\Gamma}}_{\ell\ell'} =
\nu \begin{pmatrix}
(\Clobs^{XY})^2 + \Clobs^{XX}\Clobs^{YY}
\end{pmatrix}.
\end{equation}
Forecasts presented here for cross-correlation experiments will be of this cross-only form and with noise terms given by \cref{eqn:simplenoise}.

\section{Experiments Considered}
A number of surveys across multiple wavebands are both currently taking place and planned for the near future which have weak lensing cosmology as a prominent science driver. We adopt the language of the Dark Energy Task Force \citep[DETF,][]{2006astro.ph..9591A} in loosely grouping these experiments into `Stage III' and `Stage IV' experiments, where Stage III refers to experiments which were in the near future when the DETF document was prepared compared to Stage IV experiments which follow these in time. The distinction can also be cast in terms of the expected level of constraining power, with Stage III Weak Lensing alone experiments giving $\mathcal{O}(50\%)$ constraints on the Dark Energy equation of state $w$ and Stage IV $\mathcal{O}(10\%)$. We point out that we present here constraints from weak lensing analyses only; in reality, significant improvements on constraints will be gained by both the SKA and optical surveys' measurements of galaxy clustering and other probes (such as supernovae and Intensity Mapping), as well as combination with external data sets.

For each stage we consider a representative experiment from both the optical and the radio. We now give short background descriptions of the source populations assumed and the particulars of each experiment considered.
\label{sec:experiments}
\begin{table*}
\begin{tabular}{lccccccccccc}
\hline 
Experiment & $A_{\rm sky} \, [\mathrm{deg}^2]$ & $n_{\rm gal} \, [\mathrm{arcmin}^{-2}]$ & $z_{m}$ & $\alpha$ & $\beta$ & $\gamma$ & $f_{\textrm{spec-}z}$ & $z_{\textrm{spec-max}}$ & $\sigma_{\textrm{photo-}z}$ & $z_{\textrm{photo-max}}$ & $\sigma_{\textrm{no-}z}$ \\
\hline
SKA1 & 5,000 & 2.7 & 1.1 & $\sqrt{2}$ & 2 & 1.25 & 0.15 & 0.6 & 0.05 & 2.0 & 0.3 \\
\des & 5,000 & 12 & 0.6 & $\sqrt{2}$ & 2 & 1.5 & 0.0 & 2.0 & 0.05 & 2.0 & 0.3 \\
\hline
SKA2 & 30,000 & 10 & 1.3 & $\sqrt{2}$ & 2 & 1.25 & 0.5 & 2.0 & 0.03 & 2.0 & 0.3 \\
\euclid & 15,000 & 30 & 0.9 & $\sqrt{2}$ & 2 & 1.5 & 0.0 & 0.0 & 0.03 & 4.0 & 0.3 \\
\hline
\end{tabular}
\caption{Parameters used in the creation of simulated data sets for the representative experiments considered in this paper.}
\label{tab:experiments}
\end{table*}
\subsection{Source Populations}
\begin{figure}
\includegraphics[width=0.5\textwidth]{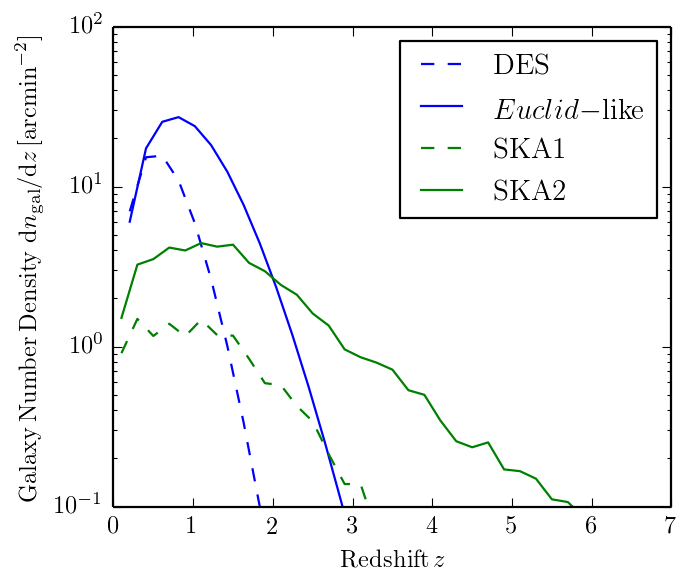}\\
\includegraphics[width=0.5\textwidth]{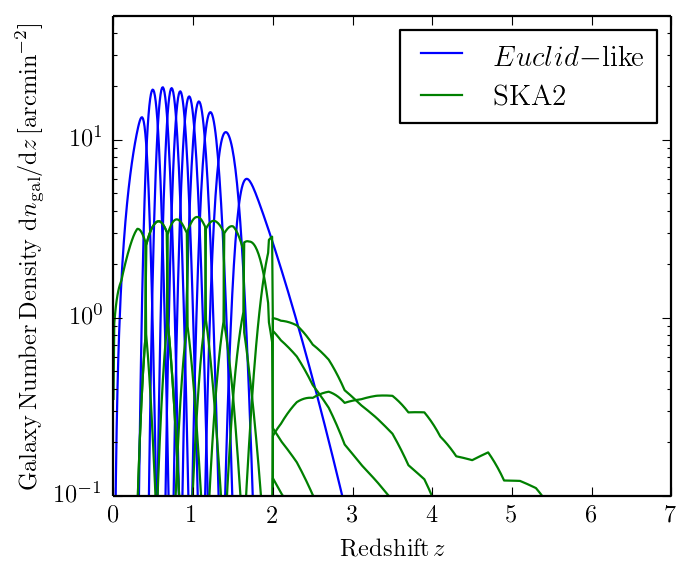}
\caption{Source (top) and ``observed" (bottom, split into ten tomographic bins for each experiment) redshift distributions $\d n_{\rm gal}/\d z$ for the \euclid and SKA2 experiments described in Section~\ref{sec:stage4_expts}. The curves in both panels are normalised such that the total area under the curves is equal to the total $n_{\rm gal}$ for each experiment.}
\label{fig:nofz}
\end{figure}

For the number density of sources in each tomographic bin in each experiment we use a redshift number density distribution of the form:
\begin{equation}
\label{eqn:nofz}
\frac{\d n_{\rm gal}}{\d z} = z^{\beta} \exp\left( -(z/z_0)^{\gamma} \right),
\end{equation}
where $z_0 = z_{m} / \alpha$ ($\alpha$ is a scale parameter) and $z_{m}$ is the median redshift of sources. For the SKA experiments we use the source counts in the SKADS S3-SEX simulation of radio source populations \citep{wilman08}; we have applied re-scalings of these populations in both size distributions and number counts in order to match recent data (see Paper II for a full description). Values of the parameters in \cref{eqn:nofz} are given in \cref{tab:experiments}, including the best-fit parameters to the SKADS S3-SEX distribution. The top panel of \cref{fig:nofz} shows these distributions for the experiments considered, including the high-redshift tail present in the radio source populations. For each experiment we then subdivide these populations into ten tomographic redshift bins, giving equal numbers of galaxies in each bin. We also add redshift errors, spreading the edges of each redshift bin and causing them to overlap. We assume a fraction of sources with spectroscopic redshifts (i.e. with no redshift error) $f_{\textrm{spec-}z}$ up to a redshift of $z_{\textrm{spec-max}}$. For the remaining sources we assign a Gaussian-distributed (with the prior $z > 0$)  redshift error of width $(1+z)\sigma_{\textrm{photo-}z}$ up to a redshift of $z_{\textrm{photo-max}}$, beyond which we assume no `good' photometric redshift estimate and assign a far greater error $(1+z)\sigma_{\textrm{no-}z}$. Values for these parameters for each representative experiment are shown in \cref{tab:experiments} and the resulting binned distributions for SKA2 and the \euclid experiment (see Section~\ref{sec:stage4_expts} below) are shown in the lower panel of \cref{fig:nofz}. We take an intrinsic galaxy shape dispersion of $\sigma_{g_i} = 0.3$ for all redshift bins and experiments\tbf{, consistent with that found for the radio and optical lensing samples used in previous radio weak lensing \citep{2010MNRAS.401.2572P}}.
\subsection{Stage III Experiments}
\subsubsection{SKA Phase 1 (SKA1)}
The Square Kilometre Array (SKA) will be built in two phases: the first (SKA1) will consist of a low frequency aperture array in Western Australia (SKA1-LOW) and a dish array to be built in South Africa (SKA1-MID) with expected commencement of science observations in 2020. Of these, it is SKA1-MID which will provide the necessary sensitivity and resolution to conduct weak lensing surveys. Here we have assumed source number densities expected to emanate from a $5,000 \, \mathrm{deg}^2$ survey conducted at the centre of observing Band 2 (1.355 GHz) and with baselines weighted to give an image-plane PSF of size 0.5 arcsec full width at half maximum (FWHM). This experimental configuration is expected to give a close-to-optimal combination of high galaxy number density and quiescent ionosphere, as well as maximise commensality with other SKA science goals (see Paper II and \citealt{2015arXiv150706639H} for further discussion). We then calculate the expected sensitivity of the instrument when used in this configuration using the curves from the SKA1 Imaging Science Performance Memo \citep{braun2014}, which assumes a two year survey, and including all sources which are resolved and detected at a signal-to-noise greater than 10. We note that estimates for the number densities and distribution of sizes for SFGs at micro-Jansky fluxes are currently somewhat uncertain. To arrive at our estimates, we follow the procedure described in Paper II. In brief, we once again make use of the SKADS S3-SEX simulation \citep{wilman08} but we have re-calibrated the absolute numbers and sizes of SFGs found in that simulation so that they match the latest observational data from deep radio surveys. For both SKA experiments we also include fractions of spectroscopic redshifts, obtained by detection of HI line emission from the source galaxies.

\subsubsection{Dark Energy Survey (DES)}
For our Stage III optical weak lensing survey we follow the performance specifications of the weak lensing component of the Dark Energy Survey (DES). DES is an optical survey with a primary focus on weak lensing cosmology, covering $5,000 \, \mathrm{deg}^2$ of the Southern hemisphere sky using the 4-metre Blanco telescope at the Cerro Tololo Inter-American Observatory in Chile. It has already produced cosmological parameter measurements from weak lensing with Science Verification data \citep{2015arXiv150705552T} and represents a `Stage III' weak lensing survey along with contemporaries such as the Kilo-Degree Survey \citep[KiDS,][]{2015MNRAS.454.3500K} and Hyper Suprime Cam (HSC) weak lensing projects. Here we use the expected performance of the full five year survey data, with observations in $g,r,i,z,Y$ bands and a limiting magnitude of 24. \tbf{The achievable weak lensing source number densities and redshift distributions considered here are drawn from \citep{2005astro.ph.10346T,2016MNRAS.tmp..452D}}.

\subsection{Stage IV Experiments}
\label{sec:stage4_expts}
\subsubsection{Full SKA (SKA2)}
\tbf{As described in \cite{dewdney2013}, the full SKA (SKA2)} will be a significant expansion of SKA1, with the current plan for SKA-MID increasing the number of dishes from 194 to $\sim 2000$ (with the initial 194 integrated into the larger array) and spreading long baselines over Southern Africa, undergoing construction between 2023 and 2030. As the sensitivity scales with approximately the total collecting area, for SKA2 we assume a ten times increase in sensitivity of the instrument and make our forecasts for a $3\pi$ steradian survey, again at the centre of observing Band 2 (1.355 GHz) and with a 0.5 arcsec PSF.
\subsubsection{\euclid}
For a Stage IV optical weak lensing experiment we consider as a reference a space-based survey capable of obtaining a galaxy number density of $n_{\rm gal} = 30 \, \mathrm{arcmin}^2$ over $15,000\,\,\mathrm{deg}^2$ of the sky, with more accurate photometric redshifts than the \des survey, but still no spectroscopic redshift measurements. We expect this to be similar to the performance of the weak lensing component of the \textit{Euclid} satellite \citep{Laureijs:2011gra,Amendola:2012ys} planned for launch in 2020. We refer to this representative Stage IV optical weak lensing-only experiment as ``\euclid".
\subsection{Cross-Correlations}
\label{sec:cross-correlations}
For cross-correlation experiments, we take combinations of Stage III experiments (\des and SKA1) and Stage IV experiments (\euclid\ and SKA2). For \des$\times$SKA1 we assume the $5,000 \degsq$ sky coverage is the same for both surveys and construct theoretical power spectra $C_{\ell}$ with lensing kernels given by $g^{\mathrm{\des}_i}$ and $g^{\mathrm{SKA1}_i}$, with ten tomographic bins from each experiment defined to have equal numbers of sources in each bin (i.e. bin $i$ for \des does not correspond to, but may overlap with, bin $i$ for SKA1). For the noise power spectra $\mathcal{N}_\ell^{X_i Y_j}$ we assume a limiting case in which there is negligible overlap between the source populations probed by the different experiments \citep[as found in][]{2015arXiv150705977D} and for objects which do exist in both surveys, shapes are uncorrelated, as suggested by the findings of \cite{2010MNRAS.401.2572P}, meaning the populations in the twenty different bins are treated as wholly independent. \tbf{As demonstrated in \cref{fig:FoM_fOR}, the relaxation of this assumption should not significantly affect the achievable constraints. In the case where the samples are completely separate, redshift information will be necessary for the SKA sources, but could be obtainable from sub-threshold techniques which make use of the HI 21 cm line below the detection limit traditionally used for spectroscopic redshifts (techniques we are exploring in ongoing work), something which should be very capable in providing imperfect $\d n_{\rm gal}/\d z$ (in the manner of photometric redshifts) for tomographically binned sources.}

For {\euclid}$\times$SKA2 we consider only the $15,000 \degsq$ survey region available to both experiments. Again, ten equally populated tomographic redshift bins are chosen for each experiment and observed cross-spectra are formed. We emphasise that we are not merely considering the lowest $n_{\rm gal}$ of the two experiments for the cross-correlations, but using the full $\d n_{\rm gal}/\d z$ distributions in twenty bins, ten from each experiment, making use of all the galaxies present.

\section{Forecasting Methods}
\label{sec:forecasts}
For forecasting constraints on cosmological parameters which will be possible with the SKA and cross-correlations we use two approaches: Markov Chain Monte Carlo (MCMC) mapping of the likelihood distribution and the Fisher Matrix approximation. For a given likelihood function and covariance matrix, MCMC methods are accurate and capable of tracing complicated posterior probability distribution surfaces in multiple dimensions, but are computationally expensive. Here, we run MCMC chains for all of our experiments and use them as a calibration for Fisher matrices, allowing the latter to be robustly used for future similar work. The calculation of realistic covariance matrices beyond the approximation in \cref{eqn:simple_cov} typically requires large-scale simulations of data of the type expected to be generated in an experiment; in Paper II we construct such simulations for a fiducial cosmology.
\subsection{Forecasts with \cosmosis}
For our MCMC parameter constraint forecasts we make use of the \cosmosis modular cosmological parameter estimation code \citep{2015A&C....12...45Z}. For a given set of cosmological parameters $\boldsymbol\vartheta$ we calculate a non-linear matter power spectrum using CAMB \citep{2000ApJ...538..473L} (with modifications from ISiTGR for the modified gravity models from \citealt{2011PhRvD..84l3001D, 2015PhRvD..92b3003D}) and halofit \citep{2003MNRAS.341.1311S, 2012ApJ...761..152T}. This is then converted to a shear power spectrum using \cref{eqn:limber} and the assumed $n_{X_i}(z)$ for the relevant experiment and redshift bin.

These shear power spectra are compared in a Gaussian likelihood to an `observed' data vector $\widetilde{d}_{\ell}$ and covariance matrix, calculated using the same method at our fiducial cosmological parameters:
\begin{align}
-2 \ln \mathcal{L} &= \nonumber \\
& \sum_{\ell,\ell'=\ell_\mathrm{min}}^{\ell_\mathrm{max}} \left(C_\ell^{XY}(\boldsymbol\vartheta) - \widetilde{d}_{\ell}\right) \left[ \mathbf{\Gamma}^{XY}_{\ell\ell'} \right]^{-1} \left(C_{\ell'}^{XY}(\boldsymbol\vartheta) - \widetilde{d}_{\ell'}\right),
\end{align}
summing over all multipoles as $\Gamma^{XY}_{\ell\ell'}$ is assumed to be diagonal in $\ell$ and $\ell'$. We then use the MultiNest \citep{2013arXiv1306.2144F} code to sample over this parameter space and form the posterior confidence regions shown in our results plots. For all of our MCMC forecasts we include information up to a multipole of $\ell_{\rm max} = 3000$, capturing mildly non-linear scales, dependent on the redshift being probed.
\subsection{Comparison with Fisher Matrices}
\label{sec:fisher}
Whilst fully sampling the posterior distribution with Markov Chain methods provides a robust and accurate prediction for parameter constraints, it is typically computationally expensive and time consuming.
The Fisher matrix is an alternative approach for parameter estimation which assumes the presence of
a likelihood function $L(\boldsymbol\vartheta)$ that quantifies the
agreement between a certain set of experimental data and the set of parameters of
the model, $\boldsymbol\vartheta=\{\vartheta_\alpha\}$. It also assumes that
the behaviour of the likelihood near its maximum characterises the whole 
likelihood function sufficiently well to be used to estimate errors on the 
model parameters \citep{Jeffreys:1961,1996ApJ...465...34V,Tegmark:1996bz}. 

Under the hypothesis of a Gaussian likelihood, the Fisher matrix 
is defined as the inverse of the parameter covariance matrix. Thence, it is
possible to infer the statistical accuracy with which the data encoded in the 
likelihood can measure the model parameters. If the data is
taken to be the expected measurements performed by future experiments, the Fisher 
matrix method can be used, as we do here, to determine its prospects for 
detection and the corresponding level of accuracy. The $1\sigma$ marginal error on 
parameter $\vartheta_\alpha$ reads
\begin{equation}
\sigma(\vartheta_\alpha) = \sqrt{ \left( \mathbfss F^{-1} \right)_{\alpha\alpha}},
\label{eq:marginal}
\end{equation}
where $\mathbfss F^{-1}$ is the inverse of the Fisher matrix, and no summation 
over equal indices is applied here.

Our experimental data will come from the measurement of the (cross-)correlation 
angular power spectrum $C^{XY}_\ell$ between the observables $X$ and $Y$. From an observational point of view, we can consider each single mode 
$\widetilde{C}^{XY}_\ell$ in tomographic and multipole space as a parameter of the theory. Then, to recast the Fisher matrix in the space of the model parameters, $\boldsymbol\vartheta$, it is sufficient to multiply the inverse of the covariance matrix by the Jacobian of the change of variables, viz.\
\begin{equation}
\mathbfss{F}_{\alpha\beta} = \sum_{\ell,\ell'=\ell_\mathrm{min}}^{\ell_\mathrm{max}}
\frac{\partial \mathbfss{C}^{XY}_\ell}{\partial \vartheta_\alpha}
\left[ \mathbf{\Gamma}^{XY}_{\ell\ell'} \right]^{-1}
\frac{\partial \mathbfss{C}^{XY}_{\ell'}}{\partial \vartheta_\beta},
\label{eq:fisher}
\end{equation}
where again we sum over all the multipoles because $\Gamma^{XY}_{\ell\ell'}$ is here assumed to be diagonal in $\ell$ and $\ell'$.

\begin{figure}
\includegraphics[width=0.5\textwidth]{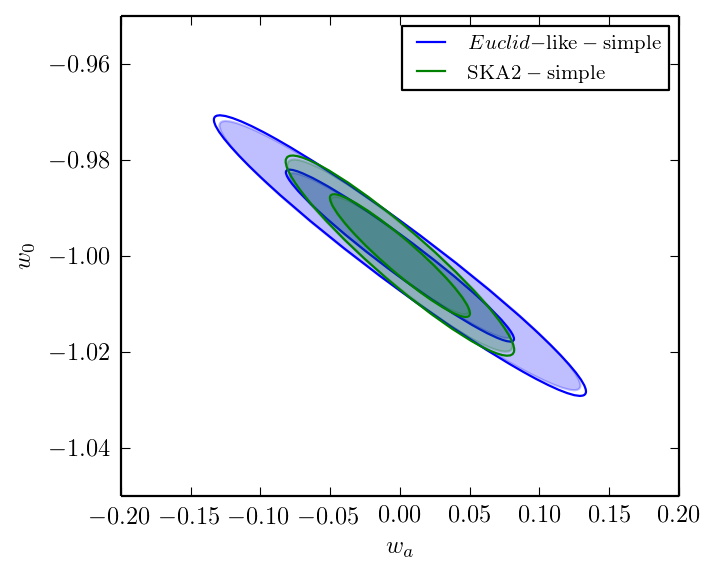}
\caption{Fisher (unfilled contours) and MCMC (filled contours) predictions for the simplified weak lensing-only experiments considered in \cref{sec:fisher}, showing agreement in both size and degeneracy direction. One dimensional uncertainties for both cases are shown in \cref{tab:fisher_comparison}.}
\label{fig:fisher_comp}
\end{figure}

Fisher matrices can be quickly computed, requiring computation of observational shear spectra only at the set of points in parameter space necessary for approximating the derivative, rather than at enough points to create a good, smooth approximation to the true posterior. This allows exploration of the impact of different systematics and analysis choices on forecast parameter constraints, which we intend to explore in a following paper. Here, we validate the use of the Fisher approximation for such an exploration by comparing for a simple case the predictions from our MCMC chains and Fisher matrices. We use simplified versions of the SKA2 and \euclid experiments (intended to maximise the Gaussianity of the contours and be quicker to compute), in which we consider both as covering the full sky ($A_{\rm sky} = 41,253 \, \degsq$), only use information up to $\ell=1000$ and cut off both redshift distributions at $z=4$. For these simplified experiments we calculate the parameter covariance matrix in the two parameters  $\lbrace w_0, w_a \rbrace$ using both the MCMC procedure and via the Fisher matrix approximation. \Cref{fig:fisher_comp} shows confidence region ellipses corresponding to both these methods and \cref{tab:fisher_comparison} the associated one dimensional parameter constraints, showing $\mathcal{O}(5\%)$ agreement.

\tbf{As a demonstration of the usefulness of this approach, we show the benefit of the high-redshift tail in the source distribution for SKA by calculating constraints in $\lbrace w_0, w_a \rbrace$ both including and excluding all sources above $z=2$. For SKA1, excluding these sources leads to a $\lbrace 3.63, 4.04 \rbrace$ factor increase in the width of the uncertainties, whilst for SKA2 the factors are $\lbrace 1.32, 1.51 \rbrace$.}
\begin{table}
\centering
\begin{tabular}{lcc}
\hline 
Experiment & $\sigma_{\w}$ MC, Fisher & $\sigma_{\wa}$ MC, Fisher \\
\hline
SKA2-simple & 0.0161, 0.0168 & 0.0651, 0.0660 \\
\euclid-simple & 0.0226, 0.0236 & 0.104, 0.108 \\
\hline
\end{tabular}
\caption{One dimensional parameter constraints from covariance matrices calculated using full MCMC chains and the Fisher matrix formalism for the simplified weak lensing-only experiments described in \cref{sec:fisher}, showing good agreement, as shown in \cref{fig:fisher_comp}. The constraints for SKA2 correspond to a DETF figure-of-merit of $\sim2500$.}
\label{tab:fisher_comparison}
\end{table}

\section{Results}
\label{sec:results}
\begin{figure*}
\includegraphics[width=0.475\textwidth]{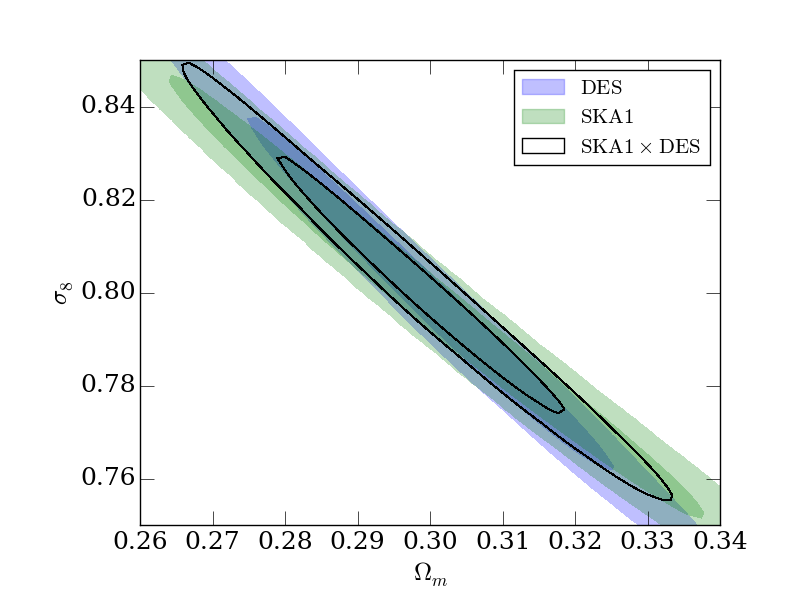}\includegraphics[width=0.475\textwidth]{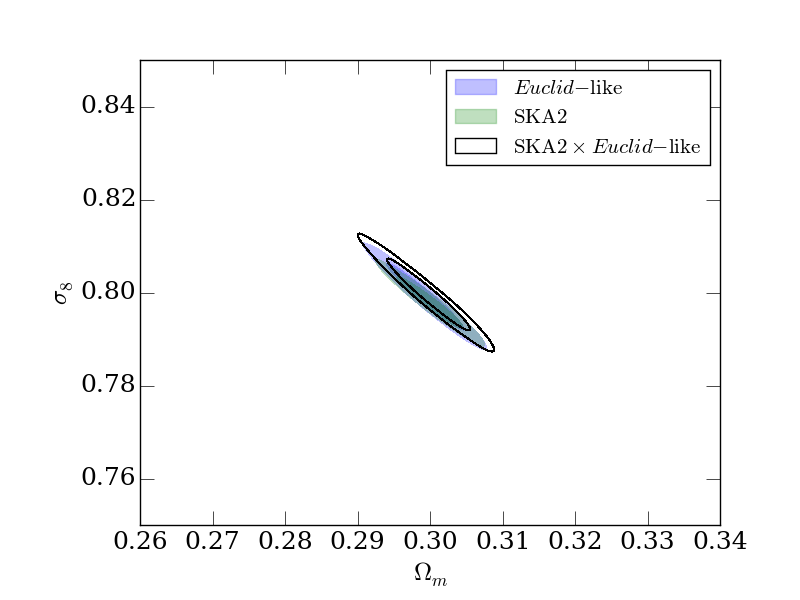}
\caption{Stage III (left) and Stage IV (right) weak lensing-only constraints on matter content ($\sigma_8$,$\om$) parameters, including those from cross-correlation spectra between SKA1 and DES, and between SKA2 and the \euclid experiment.}
\label{fig:matter}
%\end{figure*}
%
%\begin{figure*}
\includegraphics[width=0.475\textwidth]{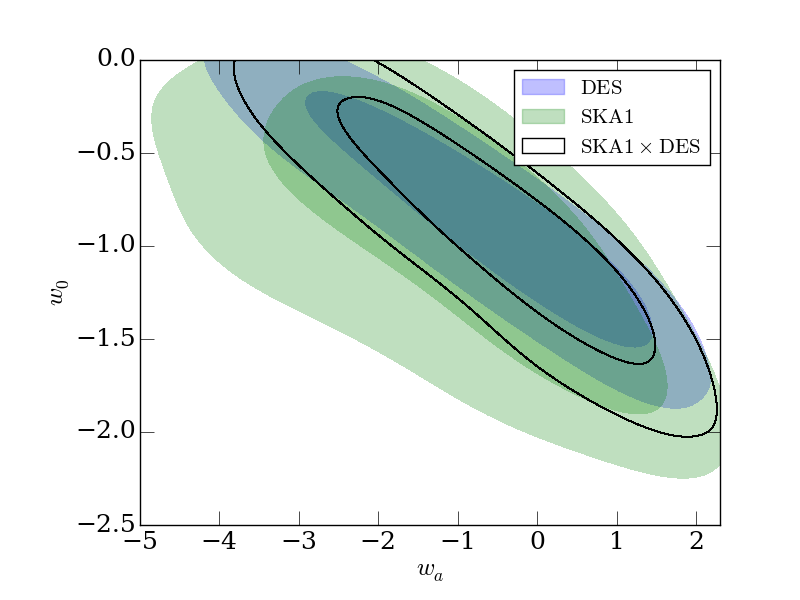}\includegraphics[width=0.475\textwidth]{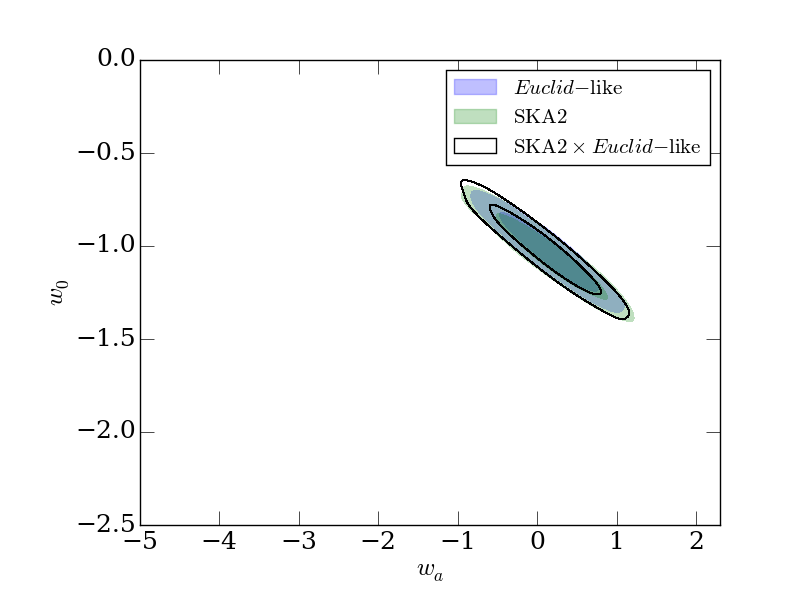}
\caption{Stage III (left) and Stage IV (right) weak lensing-only constraints on dark energy ($w_0$,$w_a$) parameters, including those from cross-correlation spectra between SKA1 and DES, and between SKA2 and the \euclid experiment. Note the different axis scales between the two plots.}
\label{fig:w0wa}
%\end{figure*}
%
%\begin{figure*}
\includegraphics[width=0.475\textwidth]{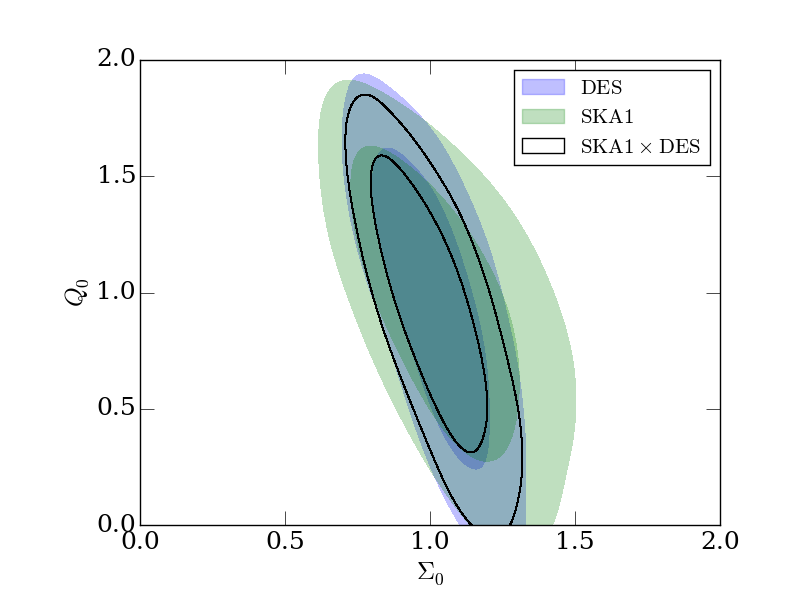}\includegraphics[width=0.475\textwidth]{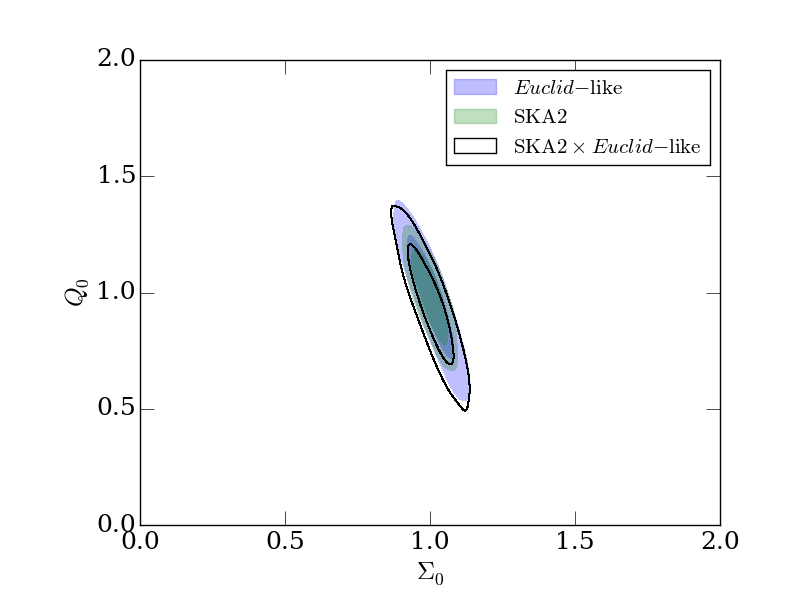}
\caption{Stage III (left) and Stage IV (right) weak lensing-only constraints on modified gravity ($\Sigma_0$,$Q_0$) parameters, including those from cross-correlation spectra between SKA1 and DES, and between SKA2 and the \euclid experiment.}
\label{fig:mg}
\end{figure*}

In \crefrange{fig:matter}{fig:mg} we show the two dimensional parameter constraints \tbf{from our MCMC forecasts} on matter $\lbrace \sigma_8, \om \rbrace$, dark energy $\lbrace w_0, w_a \rbrace$ and modified gravity $\lbrace \Sigma_0, Q_0 \rbrace$ parameter pairs, each marginalised over the full base \lcdm parameter set $\lbrace \om, \omb, \sigma_8, h_0, n_s \rbrace$, with the light (dark) regions representing $95\%$ ($68\%$) confidence regions for the parameter values, and \cref{tab:marginals} showing one dimensional 1$\sigma$ confidence regions for each parameter individually. \Cref{tab:marginals} also shows the DETF Figure of Merit (FoM) for each experiment, calculated as the inverse area of a elliptical confidence region defined from the calculated parameter covariance matrix of the simulated experiments:
\begin{equation}
\label{eqn:de_fom}
\mathrm{FoM} = \left( \sigma_{w_0} \sigma_{w_a} \sqrt{1 - \rho^2}\right)^{-1}    
\end{equation}
where $\rho$ is the correlation coefficient and $\sigma_{w_0}$ and $\sigma_{w_a}$ are the one dimensional parameter standard deviations.

The left column of \crefrange{fig:matter}{fig:mg} shows these for the three Stage III experiments: \des, SKA1 and their cross-correlation. SKA1 performs only slightly worse than \des, to be expected due to the significantly lower galaxy number density, some of which deficit is made up for by the higher-median redshift distribution, which may be expected to provide a stronger lensing signal. The DES$\times$SKA1 contours, which make use of all of the galaxies in both experiments, outperform each experiment individually in the $\lbrace \sigma_8, \om \rbrace$ case.

The right column of \crefrange{fig:matter}{fig:mg} shows the constraints for Stage IV experiments. Here, SKA2, for which Galactic foregrounds are not a consideration and hence has access to a full $30,000\,\degsq$, outperforms the \euclid experiment in the $\lbrace \sigma_8, \om \rbrace$ contours. The cross-correlation contours, which only include galaxies in the $15,000\,\degsq$ available to both experiments are slightly larger than the individual experiments, but may be expected to be significantly more robust due to the removal of wavelength-dependent systematics.
\subsection{Application of \planck Priors}
\begin{figure*}
\includegraphics[width=0.475\textwidth]{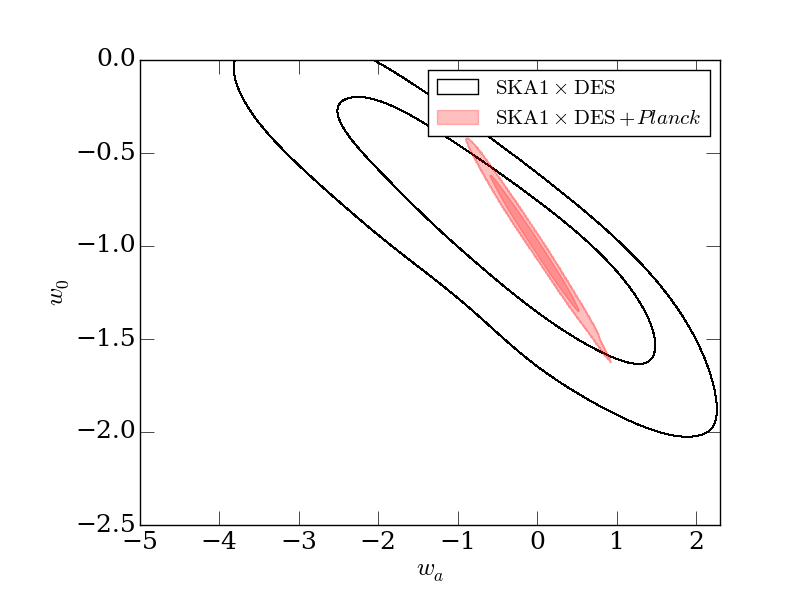}\includegraphics[width=0.475\textwidth]{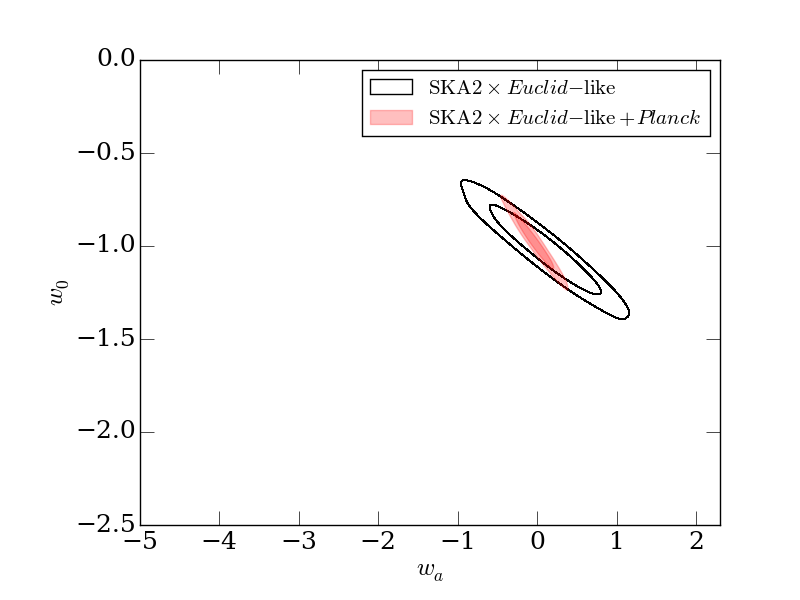}
\caption{Dark energy ($w_0$,$w_a$) parameter constraints when Stage III and Stage IV weak lensing-only experiments are combined with Cosmic Microwave Background priors from \citet{2015arXiv150201589P}.}
\label{fig:planck_priors}
\end{figure*}

We also show constraints obtained by combining the results from our experiments with results from observations of the CMB by the \planck satellite \citep{2015arXiv150201589P} in \cref{fig:planck_priors}. For this, we re-weight our MCMC chains using the plikHM-TTTEEE-lowTEB-BAO \planck likelihood chain\footnote{Obtained from the Planck Legacy Archive \url{http://www.cosmos.esa.int/web/planck/pla}}, re-centred around our fiducial cosmology. We also show the combined, marginalised parameter constraints for both auto and cross-correlation experiments in \cref{tab:marginals}. Whilst these result in little difference in the matter parameters, the different degeneracy direction of the \planck constraints on $(w_0,w_a)$ allows for a significantly smaller area in the contours, improving the DETF FoM by a factor $\sim5$ for each experiment and allowing $\mathcal{O}(10\%)$ constraints on both parameters.

\begin{table*}
\begin{tabular}{lrl|rl|rl|c}
\hline 
Experiment & ($\sigma_{\om}/\om$, & $\sigma_{\sigma_{8}}/\sigma_8$) & ($\sigma_{\w}$, & $\sigma_{\wa}$) & ($\sigma_{\Sigma_{0}}/\Sigma_0$, & $\sigma_{Q_{0}}/Q_0$) & DETF FoM\\
\hline
SKA1 & 0.083 & 0.040 & 0.52 & 1.6 & 0.19 & 0.43 & 1.6\\
SKA1 + \planck & 0.084 & 0.040 & 0.28 & 0.43 & - & - & 77\\
\des & 0.056 & 0.032 & 0.43 & 1.4 & 0.13 & 0.43 & 3.5\\
\des + \planck & 0.058 & 0.033 & 0.22 & 0.33 & - & - & 89\\
SKA1$\times$\des & 0.046 & 0.024 & 0.45 & 1.3 & 0.13 & 0.39 & 3.3\\
SKA1$\times$\des + \planck & 0.046 & 0.024 & 0.23 & 0.36 & - & - & 106\\
\hline
SKA2 & 0.010 & 0.0046 & 0.14 & 0.42 & 0.04 & 0.13 & 51\\
SKA2 + \planck & 0.010 & 0.0047 & 0.086 & 0.15 & - & - & 305\\
\euclid & 0.011 & 0.0058 & 0.13 & 0.38 & 0.053 & 0.17 & 54\\
\euclid + \planck & 0.012 & 0.059 & 0.095 & 0.16 & - & - & 244\\
SKA2$\times$\euclid & 0.013 & 0.0064 & 0.15 & 0.43 & 0.053 & 0.17 & 45\\
SKA2$\times$\euclid + \planck & 0.013 & 0.0064 & 0.10 & 0.17 & - & - & 240\\
\hline
\end{tabular}
\caption{One dimensional marginalised constraints on the parameters considered, where all pairs (indicated by brackets) are also marginalised over the base \lcdm parameter set.}
\label{tab:marginals}
\end{table*}

\section{Conclusions}
\label{sec:conclusions}
In this paper we have presented forecasts for cosmological parameter constraints from weak lensing experiments involving the Square Kilometre Array (SKA), both in isolation and in cross-correlation with comparable optical weak lensing surveys. We have shown that the first phase of the SKA (SKA1) can provide $\mathcal{O}(5\%)$ constraints on matter parameters $\om$ and $\sigma_8$, $\mathcal{O}(50\%)$ constraints on dark energy equation of state parameters $w_0$ and $w_a$, and $\mathcal{O}(10\%)$ constraints on modified gravity parameters $\Sigma_0$ and $Q_0$, competitive with the Dark Energy Survey (DES). The full SKA (SKA2) can significantly improve on all of these constraints and be competitive with the surveys planned with Stage IV optical weak lensing experiments. Furthermore, we have explored what may be achieved with weak lensing constraints from the cross-correlation power spectra between radio and optical experiments. Such cross-correlation experiments are important as they will be free of wavelength-dependent systematics which can otherwise cause large biases which dominate statistical errors and can lead to erroneous cosmological model selection. For both the Stage III (SKA1, \des) and Stage IV (SKA2, \euclid) experiments, such systematics are potentially larger than the statistical errors available from the number density of galaxies probed. We have shown that parameter constraints made using only the cross-waveband power spectra can be as powerful as traditional approaches considering each experiment separately, but with the advantage of being more robust to systematics. Such cross-correlation experiments represent significant promise in allowing weak lensing to maximise its potential in extracting cosmological information. At both Stage III and Stage IV, constraints on $(w_0,w_a)$ are significantly improved with the addition of Cosmic Microwave Background priors from the \planck satellite, down to $\mathcal{O}(10\%)$ in both parameters for SKA2 + \planck.

The realisation of this promise in practice will rely on a number of developments:
\begin{itemize}
\item The accuracy and reliability of shape measurements of galaxies from SKA data (which will arrive in the poorly-sampled Fourier plane as visibilities) will need to match that available from image-plane optical experiments \citep[see][for further discussion]{2015aska.confE..30P}.
\item Understanding of the star-forming radio galaxy populations making up the sources in SKA weak lensing surveys, and how these correspond to the source populations in optical surveys.
\tbf{\item The extraction of redshift information for the radio sources, either from cross-matching catalogues, requiring deep data in wavebands capable of providing photometric redshifts, or extracting HI 21 cm line redshifts from below a traditional survey threshold.}
\item Optimisation of SKA survey strategies to maximise the amount of information gained in radio weak lensing surveys. For more discussion of this see \cite{bonaldi2016} (Paper II).
\item Inclusion of additional information from radio polarisation and spectral line measurements, which may mitigate other, wavelength-independent systematics which are not removed by cross-correlations, such as galaxy intrinsic alignments. We intend to explore the impact of these approaches on parameter constraints in a future work using Fisher matrix forecasts to quantify the impact of such systematics and how well they may be removed.
\end{itemize}

These problems are currently being addressed, through the radioGREAT data simulation programme\footnote{\url{http://radiogreat.jb.man.ac.uk}}, precursor experiments and exploitation of archival data \cite[][SuperCLASS]{2015arXiv150705977D}, large scale simulations (Paper II) and theoretical work \citep[e.g.][]{2015MNRAS.451..383W}. If these aspects can be understood sufficiently well, the use of radio and radio-optical cross-correlation experiments will maximise the potential of weak lensing experiments, allowing us to more closely approach the full precision available from the data and give the best chance possible of starting to understand the true physical nature of dark matter and dark energy.

\section*{Acknowledgments}
IH, SC and MLB are supported by an ERC Starting Grant (grant no. 280127). MLB is an STFC Advanced/Halliday fellow. JZ is supported by an ERC Starting Grant (grant no. 240672). We thank Anna Bonaldi for useful discussions and Constantinos Demetroullas and Ben Tunbridge for help with the matched radio-optical catalogues.

\bibliography{ska_wl_forecasts}
\bibliographystyle{mn2e_plus_arxiv}

\label{lastpage}

\end{document}